\documentclass[proof]{pasj00}

\begin{document}
\SetRunningHead{Yamaki et al.}{Smoothed Bandpass Calibration}
\Received{2012/3/21}
\Accepted{2012/5/17}

\title{Optimization by Smoothed Bandpass Calibration in Radio Spectroscopy}

\author{Haruka \textsc{Yamaki}, Seiji \textsc{Kameno}\thanks{Corresponding author: Seiji \textsc{Kameno}}, Hirohisa \textsc{Beppu}, Izumi \textsc{Mizuno}, and Hiroshi \textsc{Imai}}%

\affil{Department of Physics and Astronomy, Graduate School of Science and Engineering, Kagoshima University, \\ 1-21-35, Korimoto, Kagoshima, 890-0065}
\email{kameno@sci.kagoshima-u.ac.jp}

%

\KeyWords{telescope---technics: calibration---techniques: spectroscopy---instrumentation: detectors---methods: statistical} 

\maketitle

\begin{abstract}
We have developed the Smoothed Bandpass Calibration (SBC) method and the best suitable scan pattern to optimize radio spectroscopic observations.
Adequate spectral smoothing is applied to the spectrum toward OFF-source blank sky adjacent to a target source direction for the purpose of bandpass correction.
Because the smoothing process reduces noise, the integration time for OFF-source scans can be reduced keeping the signal-to-noise ratio.
Since the smoothing is not applied to ON-source scans, the spectral resolution for line features is kept.
An optimal smoothing window is determined by bandpass flatness evaluated by Spectral Allan Variance (SAV).
An efficient scan pattern is designed to the OFF-source scans within the bandpass stability timescale estimated by Time-based Allan Variance (TAV).
We have tested the SBC using the digital spectrometer, VESPA, on the VERA Iriki station.
For the targeted noise level of $5 \times 10^{-4}$ as a ratio to the system noise, the optimal smoothing window was 32 -- 60 channels (ch) in the whole bandwidth of 1024 ch, and the optimal scan pattern was designed as a sequence of 70-s ON + 10-s OFF scan pairs.
The noise level with the SBC was reduced by a factor of $1.74$ compared with the conventional method.
The total telescope time to achieve the goal with the SBC was 400 s, which was $\frac{1}{3}$ of 1200 s required by the conventional way.
Improvement in telescope time efficiency with the SBC was calculated as $3\times$, $2\times$ and $1.3\times$ for single-beam, dual-beam, and on-the-fly (OTF) scans, respectively.
The SBC works to optimize scan patterns for observations from now, and also works to improve signal-to-noise ratios of archival data if ON- and OFF-source spectra are individually recorded, though the efficiency depends on the spectral stability of the receiving system.

\end{abstract}

\section{Introduction}
Spectral observations in radio astronomy are aimed to detect emission or absorption line features that bring us rich information such as intensity, velocity, line width, etc.
To acquire a spectrum, signals received by a radio telescope are processed in a spectrometer.
The performance of a spectral observation is characterized by the spectral resolution, the bandwidth, the sensitivity, and the stability. 

A certain bandpass calibration procedure is needed to obtain a desired performance of spectral observations.
An acquired spectrum is a summation of unwanted noise, $T_{\mathrm{sys}}$, and the signal from the target source that is denoted as the antenna temperature, $T_{\mathrm{a}}(\nu)$.
The spectral shape is affected by the bandpass response, $H(\nu)$, that is the transfer function of the receiving system as a function of frequency, $\nu$.
The acquired ON-source spectrum will be then $T_{\mathrm{ON}}(\nu) = H(\nu) (T_{\mathrm{sys}} + T_{\mathrm{a}}(\nu))$.
Bandpass calibration is necessary to estimate $T_{\mathrm{a}}(\nu)$ from the acquired spectrum, $T_{\mathrm{ON}}(\nu)$.

In most of radio observations, the acquired spectrum is dominated by the system noise, i.e., $T_{\mathrm{sys}} \gg T_{\mathrm{a}}(\nu)$.
To relieve the source spectrum from the system noise, a spectrum of OFF-source blank sky adjacent to the target source is subtracted from the ON-source spectrum.
This position-switching scan yields the source spectrum as
\begin{eqnarray}
T_{\mathrm{a}}(\nu) = \frac{T_{\mathrm{ON}}(\nu) - T_{\mathrm{OFF}}(\nu)}{H(\nu)}, \label{eqn:onoff_subtraction}
\end{eqnarray}
where $T_{\mathrm{OFF}}$ is the spectrum taken from the OFF-source scan.
In conventional position-switching observations, the bandpass $H(\nu)$ is also acquired from the OFF-source scans because the blank sky emits a featureless flat spectrum. For $\displaystyle H(\nu) = \frac{T_{\mathrm{OFF}}(\nu)}{T_{\mathrm{sys}}}$, equation \ref{eqn:onoff_subtraction} will be
\begin{eqnarray}
T_{\mathrm{a}}(\nu) = T_{\mathrm{sys}} \left( \frac{T_{\mathrm{ON}}(\nu)}{T_{\mathrm{OFF}}(\nu)} - 1\right). \label{eqn:onoff_subtraction2}
\end{eqnarray}

The sensitivity of spectal observations is evaluated by a SD (standard deviation), $\sigma$, of the acquired spectrum in line-free channels.
It is derived from equation \ref{eqn:onoff_subtraction2} as
\begin{eqnarray}
\left( \frac{\sigma}{T_{\mathrm{a}}(\nu) + T_{\mathrm{sys}}} \right)^2 = \left( \frac{\sigma_{\mathrm{ON}}}{T_{\mathrm{ON}}(\nu)} \right)^2 + \left( \frac{\sigma_{\mathrm{OFF}}}{T_{\mathrm{OFF}}(\nu)} \right)^2, \label{eqn:sensitivity1}
\end{eqnarray}
where $\sigma_{\mathrm{ON}}$ and $\sigma_{\mathrm{OFF}}$ are SDs of the ON- and OFF-source spectra, respectively.
For weak sources for which $T_{\mathrm{ON}}(\nu) \simeq T_{\mathrm{OFF}}(\nu)$, $\sigma$ is given by a root of sum squared (RSS) of $\sigma_{\mathrm{ON}}$ and $\sigma_{\mathrm{OFF}}$.
When $\sigma_{\mathrm{ON}}$ and $\sigma_{\mathrm{OFF}}$ are dominated by thermal noise, 
according to \citet{2009tra..book.....W}, 
they are expected to relate to the integration time, $t_{\mathrm{ON}}$ and $t_{\mathrm{OFF}}$, the spectral resolution, $\nu_{\mathrm{res}}$, and $T_{\mathrm{sys}}$ as followed,
\begin{eqnarray}
\sigma_{\mathrm{ON}} = \frac{T_{\mathrm{sys}} + T_{\mathrm{a}}}{\sqrt{\nu_{\mathrm{res}} t_{\mathrm{ON}}}}, \ \ \sigma_{\mathrm{OFF}} = \frac{T_{\mathrm{sys}}}{\sqrt{\nu_{\mathrm{res}} t_{\mathrm{OFF}}}}. \label{eqn:sigma}
\end{eqnarray}
The total telescope time is the sum of $t_{\mathrm{ON}}$, $t_{\mathrm{OFF}}$, and other overhead time such as setup, system noise measurement, gaps between scans, etc.
Optimized scan pattern is designed to maximize the sensitivity, i.e., to minimize $\sigma$, within the minimal telescope time.
Conventionally $t_{\mathrm{ON}}$ and $t_{\mathrm{OFF}}$ are equally allocated in position-switching observations to have $\sigma_{\mathrm{ON}} \simeq \sigma_{\mathrm{OFF}}$.
This manner makes the total telescope time longer than twice of $t_{\mathrm{ON}}$ and lets $\sigma \simeq \sqrt{2} \sigma_{\mathrm{ON}}$.

If there is an efficient way to reduce $\sigma_{\mathrm{OFF}}$ less than $\sigma_{\mathrm{ON}}$, we can save the total telescope time and get better sensitivity.
For an ideal receiving system, whose bandpass response were known {\it a priori} and very stable, OFF-source observations would be unnecessary.
Modern systems offer better spectral stability in the trend in replacing analog devices with a high-speed digital sampling system in an upper stream of signal transfer.
The bandpass response of a digital system, including a digital filter and a digital spectrometer, is determined by a fixed algorithm and is not affected by environmental variations.
Utility of digital devices brings us a possibility to reduce the integration time and the SD of OFF-source scans.

In this paper we propose the Smoothed Bandpass Calibration (SBC) method where the OFF-source spectrum is smoothed across the bandwidth to reduce the integration time of the OFF-source scans and the SD of the spectrum.
Section 2 describes the concept of the SBC method.
Section 3 reports test experiments that verify the SBC and that quest for the optimal parameters of a smoothing window and a scan pattern.
In section 4 we discuss the behavior of the spectral stability and estimate sensitivity improvements expected in some scan patterns.

\section{Method}\label{sec:method}
We propose relevant spectral smoothing along frequency for OFF-source scans to reduce required integration time that achieves a desired SD.
Before the spectral smoothing, a bandpass correction is employed using a template bandpass response that is presumed to be relevantly static and taken by long-time integration toward blank sky.
The bandpass-corrected OFF-source spectrum will be almost flat with slight warp due to fluctuations of the receiving system and smeared by thermal noise.
In the following subsections we formularize the spectral behavior and establish a strategy to reduce spectral unevenness.

\subsection{Description and Assumption}
As described in section 1, the purpose of OFF-source scans is to subtract $T_{\mathrm{sys}}$ and calibrate the bandpass, $H(\nu)$, which is a frequency-dependent system response.
Since the bandpass can be time-variable, we denote it as $H(\nu, t) = H_0(\nu) + H_1(\nu, t)$, where $H_0(\nu)$ is a static template bandpass and $H_1(\nu, t)$ is a fluctuation from $H_0(\nu)$.
The observed OFF-source spectrum is composed of not only $H(\nu, t)$ but also thermal noise, $\epsilon(\nu, t)$, which is assumed to be attributed to a stochastic random process following the normal distribution with the mean, $\mu = 0$, and the variance, 
$\displaystyle \sigma^2_{\mathrm{noise}} = \frac{1}{\nu_{\mathrm{res}} t_{\mathrm{integ}}}$, 
where $t_{\mathrm{integ}}$ is the integration time.
Thus the observed OFF-source spectrum will be $T_{\mathrm{OFF}}(\nu, t) = \left( H_0(\nu) + H_1(\nu, t) \right) T_{\mathrm{sys}}(1 + \epsilon(\nu, t))$, and its expectation will be $\displaystyle \left< T_{\mathrm{OFF}}(\nu, t) \right> = \left( H_0(\nu) + H_1(\nu, t) \right) T_{\mathrm{sys}}$.
Sufficiently long integration time is required for conventional OFF-source scans to reduce $\sigma_{\mathrm{noise}}$.

We assume that $H_1(\nu, t)$ is relatively smaller than $H_0(\nu)$, almost flat and smooth along the frequency, with a typical frequency scale, $\nu_{\mathrm{var}}$ of variation.
Relevant spectral smoothing for OFF-source spectra efficiently reduces $\sigma_{\mathrm{noise}}$ 
by a factor of $\displaystyle \sqrt{\frac{\nu_{\mathrm{res}}}{\nu_{\mathrm{var}}}}$
 and keeps the expectation of OFF-source spectrum.
Since the spectral smoothing will not be applied for ON-source spectra, desired spectral resolution will be kept.

\subsection{Allan Variance Decomposition}
The flatness of $H_1(\nu, t)$ is evaluated by the spectral Allan variance (SAV) analysis.
The SAV in defined as
\begin{eqnarray}
\sigma^2_{\mathrm{y}}(\Delta \nu) &=& \left< \frac{[H(\nu + \Delta \nu) - 2H(\nu) + H(\nu - \Delta \nu)]^2}{2\Delta \nu^2} \right>. \label{eqn:def_av}
\end{eqnarray}
For a small frequency span ($\Delta \nu \ll \nu_{\mathrm{var}}$) the SAV is dominated by thermal noise and follows the characteristics of $\sigma^2_{\mathrm{y}}(\Delta \nu) \propto \Delta \nu^{-2}$.
At a larger frequency span ($\Delta \nu \gg \nu_{\mathrm{var}}$), $H_1(\nu, t)$ exceeds the thermal noise and the SAV shows different characteristics from that of the thermal noise.
SAV analysis allows us to decompose $H_1(\nu, t)$ from the thermal noise.

\begin{figure}
  \begin{center}
    \FigureFile(80mm,120mm){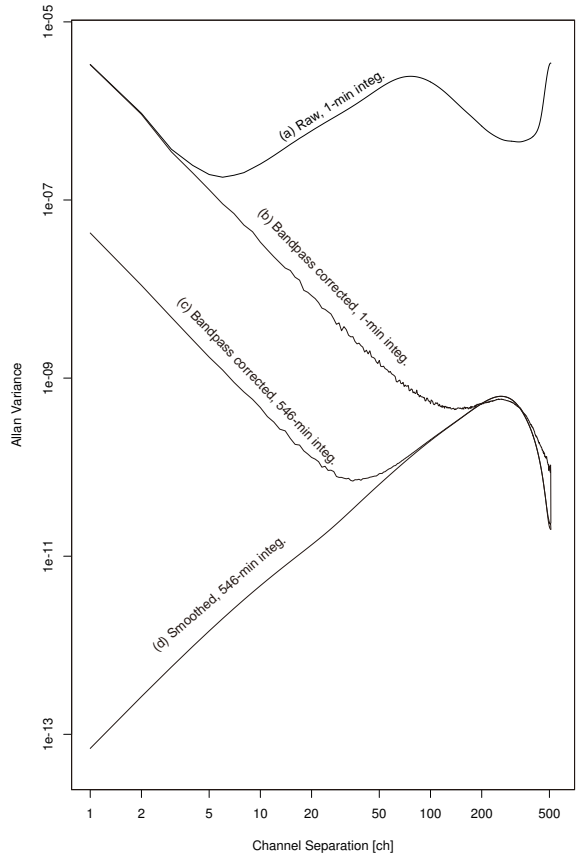}
  \end{center}
  \caption{Spectral Allan Variance (SAV) across the 16-MHz bandwidth with 1024 spectral channels.
The data comes from the field test described in section \ref{sec:tests}.
Four SAV profiles are derived from ($a$) 1-min integrated raw spectrum before bandpass correction, ($b$) 1-min integrated spectrum after bandpass correction, ($c$) 546-min integrated spectrum after bandpass correction, and ($d$) 546-min integrated spectrum after bandpass correction and spline smoothing are applied.}
  \label{fig:quantization_diagram}
\end{figure}

An example of SAV for an OFF-source spectrum is shown in figure \ref{fig:quantization_diagram},
 as a function of the channel spacing, $\nu_{\mathrm{sp}}$\footnote{As described in section \ref{sec:tests}, we used an FX spectrometer whose spectral resolution function is a squared sinc function and $\nu_{\mathrm{sp}} = \nu_{\mathrm{res}}$ by 2048-point FFT to produce 1024 ch.}.
The SAV in the 1-min integrated raw spectrum ($a$) is decomposed into two components; (1) thermal noise, $\epsilon(\nu, t)$, that dominates 
in the range of $\frac{\Delta \nu}{\nu_{\mathrm{sp}}} < 5$ and whose SAV has a dependence,
$\Delta \nu^{-2}$
 and (2) bandpass characteristics, $H(\nu, t)$, that dominates 
in the range of $\frac{\Delta \nu}{\nu_{\mathrm{sp}}} > 5$.
When components (1) and (2) show approximately symmetric power-law characteristics, the bottom of the SAV appears at the frequency span where those two components are almost equivalent (see results in subsectino 3.2 and discussion in subsection 4.1).
After bandpass correction using a template spectrum, $H_0(\nu)$, component (2) is considerably eliminated as ($b$).
The rest fluctuation, $H_1(\nu, t)$, dominates the SAV 
in $\frac{\Delta \nu}{\nu_{\mathrm{sp}}} > 120$.
As the thermal noise will be reduced via time integration in ($c$), the bottom of the SAV shifts to smaller $\Delta \nu$ 
(e.g. $\frac{\Delta \nu}{\nu_{\mathrm{sp}}} \sim 40$ for 546-min integration).
Relevant spectral smoothing with the smoothing window given by the bottom in SAV, as described in the next subsection, will yield efficient reduction of the thermal noise keeping the bandpass response as shown in ($d$).

\subsection{Strategy for Adequate Smoothing and Scans}\label{subsec:strategy}
The relevant smoothing window and ON--OFF scan pattern should be determined to minimize the SD in the OFF-source-subtracted spectrum within the given telescope time, or to minimize the total telescope time that achieves the given SD.
The best smoothing window, 
$N_{\mathrm{sw}} = \frac{\Delta \nu}{\nu_{\mathrm{res}}}$ 
is determined by the bottom of the SAV.
After we apply spectral smoothing to the OFF-source spectra, their variance is expected to be reduced to $\frac{\sigma^2_{\mathrm{OFF}}}{N_{\mathrm{sw}}}$.

Because of inequality in variances of ON- and OFF-source spectra, we need to re-design the best duty cycles of $t_{\mathrm{ON}}$ and $t_{\mathrm{OFF}}$ under the condition where total telescope time is constant.
Let an ON-to-OFF-source integration time ratio set to $1-x:x$ in each of the duty cycle and the total telescope time of $t_{\mathrm{tot}}$\footnote{Here, we omit overhead and scan gaps for the simplicity.}.
Substituting $t_{\mathrm{ON}} = (1-x) t_{\mathrm{tot}}$, $t_{\mathrm{OFF}} = x t_{\mathrm{tot}}$, and assuming that $T_{\mathrm{a}} \ll T_{\mathrm{sys}} \simeq T_{\mathrm{ON}} \simeq T_{\mathrm{OFF}}$ in equations \ref{eqn:sensitivity1} and \ref{eqn:sigma}, we have
\begin{eqnarray}
\left( \frac{\sigma}{T_{\mathrm{sys}}} \right)^2 = \frac{1}{\nu_{\mathrm{res}} t_{\mathrm{tot}}} \left( \frac{1}{1-x} + \frac{1}{N_{\mathrm{sw}} x} \right). \label{eqn:smoothed_sigma}
\end{eqnarray}
The variance is minimized to
\begin{eqnarray}
\left( \frac{\sigma_{\mathrm{min}}}{T_{\mathrm{sys}}} \right)^2 = \frac{1}{\nu_{\mathrm{res}} t_{\mathrm{tot}}} \frac{(1 + \sqrt{N_{\mathrm{sw}}})^2}{N_{\mathrm{sw}}},  \label{eqn:min_sigma}
\end{eqnarray}
when we have $\displaystyle x = \frac{1}{1 + \sqrt{N_{\mathrm{sw}}}}$.
When $N_{\mathrm{sw}} = 45$ for instance, the best ratio is $0.13$.

The interval between adjacent OFF-source scans should be shorter than the timescale of spectral stability in the receiving system.
The spectral stability is evaluated by the time-based Allan Variance (TAV).
As we'll see in figure \ref{fig:caledSpec}, a typical spectral shape of $H_1(\nu, t)$ show an {\cal S}-shaped feature with a peak and a bottom.
Since we focus on stability of the bandpass shape, the TAV is calculated by the difference between levels of the peak and the bottom, i.e. $\Delta H_1(t) = H_1(\nu_{\mathrm{peak}}, t) - H_1(\nu_{\mathrm{bottom}}, t)$. Thus, the TAV is derived as
\begin{eqnarray}
\sigma^2_{\mathrm{y}}(\tau) &=& \left< \frac{[ \Delta H_1(t + \tau) - 2 \Delta H_1(t) + \Delta H_1(t - \tau)]^2}{2\tau^2} \right>, \label{eqn:def_time_av}
\end{eqnarray}
where $\tau$ is the time lag.
The stability timescale is determined by the time lag where the TAV follows $\sigma^2_{\mathrm{y}}(\tau) \propto \tau^{-2}$.

In summary, our strategy to determine the optimal spectral smoothing window and the scan pattern will be:

\begin{enumerate}
\item
Acquire the template bandpass response, $H_0(\nu)$, and its dependence on a frequency band by sufficiently long integration before programed observations (e.g. during seasonal maintenance).
Relevant spectral smoothing can be applied to the template bandpass to reduce the thermal noise in it.

\item
Set the target SD for the observation and estimate approximate integration time.

\item Acquire test scan data to take SAV and TAV to decide $N_{\mathrm{sw}}$ at the bottom of the SAV.

\item Derive the ON--OFF duty cycle of $1-x : x$ basing on $N_{\mathrm{sw}}$.

\item Estimate the timescale of stability and determine the OFF-source-scan interval.

\item Design the optimal ON--OFF scan pattern basing on the time stability of the bandpass and the ON-OFF duty cycle.

\item Apply bandpass calibration of $H_0(\nu)$ for observed ON- and OFF-source spectra before integration. Employ spectral smoothing for the OFF-source spectra with the smoothing window of $N_{\mathrm{sw}}$.

\item Apply the bandpass calibration of $H_1$ for ON-source spectra using the adjacent smoothed OFF-source spectra. Time-integrate the bandpass-calibrated ON-source spectra.

\item Apply 
baseline subtraction, 
if necessary, to the OFF-source-subtracted spectrum and get the final result.

\end{enumerate}

\section{Tests and Results} \label{sec:tests}
Field tests have been carried out for the SBC to investigate how does it work efficiently.
We used the VERA \citep{2003ASPC..306..367K} Iriki 20-m antenna in the single-dish mode.
The antenna pointed to the zenith throughout our tests.
The 22-GHz HEMT receiver was tuned at 22.235 GHz with the bandwidth of 512 MHz.
The received signal was downconverted with the first LO of 16.8 GHz, and the second LO of 5179 MHz before digitized at 1024 Msps 2-bit quantization by the digital sampler ADS-1000 \citep{2001ExA....11...57N}.
Digital filtering was applied using the digital filter unit \citep{2005PASJ...57..259I} to split the signal into sixteen 16-MHz streams where we used only the first stream.
Spectroscopy was employed using the software spectrometer, VESPA (VEra SPectrum Analyzer; \cite{VCON2011}), that produces 
1024 ch by 2048-point FFT. 
This produces a squared sinc spectral resolution function whose first null appears at the adjacent ch, i.e., $\nu_{\mathrm{sp}} = \nu_{\mathrm{res}}$.

The template bandpass, $H_0(\nu)$, was obtained on 2010 Apr. 15 for 230-min integration.
A 10-hour observation was conducted on 2010 Oct. 5, half-a-year later than acquiring the template bandpass to include seasonal variation.
Pseudo ON--OFF scan pattern was produced from the 10-hour continuous observation; OFF-source scans was pilfered from the continuous zenith observations.
Data quality was checked via monitoring the total power of every spectrum.
The first 50-min data were not used because of unexpected operations of hot-load insertion.
We also flagged out 3-sec data because of obviously irregular spectra probably due to data transmission error.
Finally, we obtained 32768-sec continuous spectra.
For calculation of SAV, we did not use the first spectral channel that includes direct current (DC) component affected by voltage bias offset at the digital sampler.

In this test, we set the target SD of the antenna temperature of $50$ mK, which is the standard of regular monitoring program in the VERA Iriki single-dish observations.
This corresponds to $\frac{\sigma}{T_{\mathrm{sys}}} = 5 \times 10^{-4}$ for $T_{\mathrm{sys}} = 100$ K.

We employed 3rd-order B-spline smoothing for OFF-source spectra in step 7 described in subsection \ref{subsec:strategy}.
Node intervals are set to be equal to the smoothing window, $N_{\mathrm{sw}}$.
The statistical package R was used for data analyses of integration, bandpass calibration, AV calculation, and B-spline smoothing.
All of source codes for these procedures are presented in the GitHub\footnote{https://github.com/kamenoseiji/BPsmooth}.

\subsection{Bandpass Spectra}
\begin{figure}
  \begin{center}
    \FigureFile(80mm,60mm){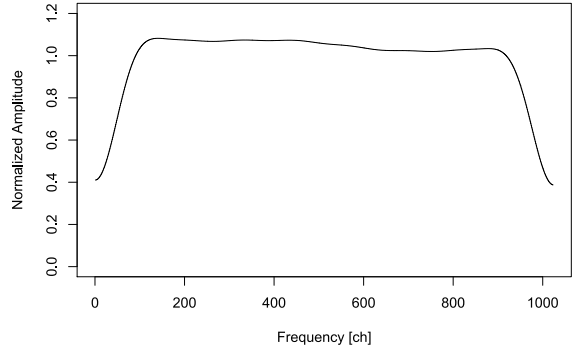}
  \end{center}
  \caption{Template bandpass spectrum, $H_0(\nu)$, obtained by 230-min integration on 2010 Apr. 15. The DC bias (spectrum at ch$=1$) is flagged out. The amplitude is normalized as the mean becomes unity.}
  \label{fig:BP}
\end{figure}

The template bandpass, $H_0(\nu)$ is shown in figure \ref{fig:BP}.
The SD of its random noise was $1.97 \times 10^{-5}$, which was estimated by $H_0(\nu) / \bar{H}_0(\nu) - 1$ where $\bar{H}_0(\nu)$ was a B-spline-smoothed spectrum.
We used $\bar{H}_0(\nu)$ for the following tests, though the SD was one order of magnitude smaller than the target noise level, to quest the ultimate performance.

\begin{figure}
  \begin{center}
    \FigureFile(80mm,60mm){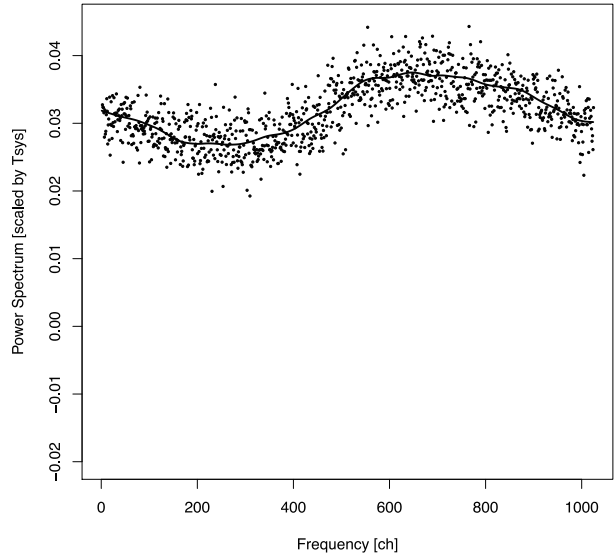}
  \end{center}
  \caption{\bf Bandpass-corrected spectrum, $\displaystyle \frac{T_{\mathrm{OFF}}(\nu, t)}{\bar{T}_{\mathrm{OFF}} \bar{H}_0(\nu)} - 1 = \frac{H_1(\nu, t)}{\bar{H}_0(\nu)}$ for 10-s average. The solid line shows the smoothed spectrum, $\displaystyle \frac{\bar{H}_1(\nu, t)}{\bar{H}_0(\nu)}$ with the smoothing window of 45 ch.}
  \label{fig:caledSpec}
\end{figure}

Bandpass-corrected spectra, $\displaystyle \frac{T_{\mathrm{OFF}}(\nu, t)}{\bar{T}_{\mathrm{OFF}} \bar{H}_0(\nu)} - 1$, are shown in figure \ref{fig:caledSpec}.
Here, $\bar{T}_{\mathrm{OFF}}$ is the mean value of the OFF-source spectra that represents $T_{\mathrm{sys}}$.
Thus the bandpass-corrected spectrum indicates $\displaystyle \frac{H_1(\nu, t)}{\bar{H}_0(\nu)} (1 + \epsilon(\nu, t)) + \epsilon(\nu, t)$.
The solid line in figure \ref{fig:caledSpec} was computed by the 3rd-order B-spline smoothing with the node intervals, $N_{\mathrm{sw}} = 45$ and represent $\displaystyle \frac{\bar{H}_1(\nu, t)}{\bar{H}_0(\nu)}$.
Departures from the solid line were dominated by the thermal noise, $\displaystyle \left( 1+ \frac{H_1(\nu, t)}{H_0(\nu)} \right) \epsilon(\nu, t) \simeq \epsilon(\nu, t)$.
The smoothed spectrum was time variable. Its time variability is described in subsection \ref{subsec:timeVariability}.

\subsection{Spectral Allan Variance}\label{subsec:SAV}
\begin{figure}
  \begin{center}
    \FigureFile(80mm,60mm){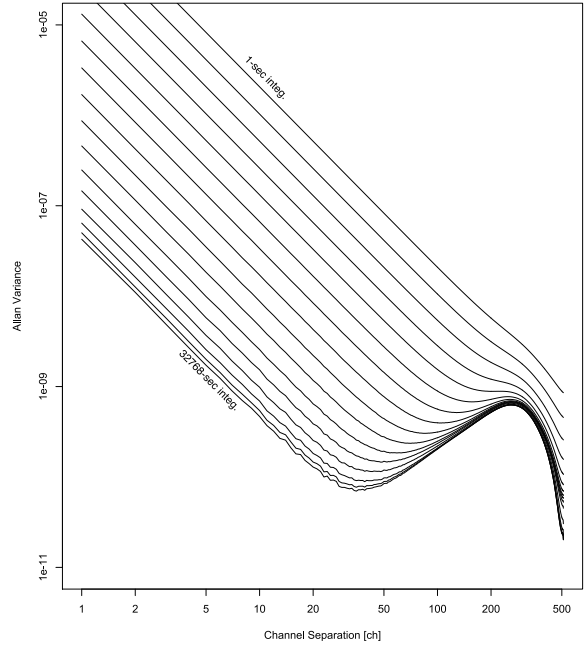}
  \end{center}
  \caption{SAVs across the bandwidth under time integration of 1, 2, 4, $\dots$, 32768 s.}
  \label{fig:spactralAV}
\end{figure}

\begin{figure}
  \begin{center}
    \FigureFile(80mm,60mm){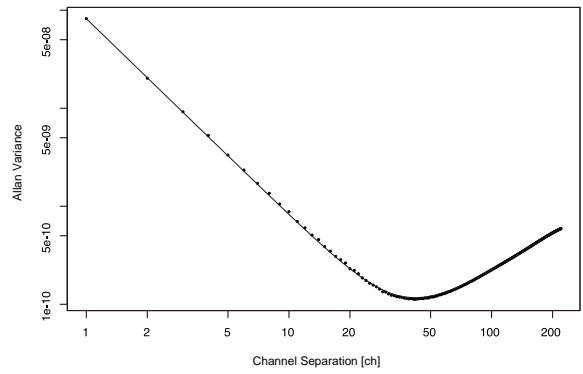}
  \end{center}
  \caption{SAV at 64-min integration. The solid line indicates the best fit with two power-law components; $\sigma^2_{\mathrm{y}}(\nu) = a_1 \nu^{\alpha_1} + a_2 \nu^{\alpha_2}$. The best-fit power indices are $\alpha_1 = -2.003 \pm 0.001$ and $\alpha_2 = 1.29 \pm 0.02$.}
  \label{fig:SAV_fit}
\end{figure}

The SAVs of bandpass-corrected spectra defined as equation \ref{eqn:def_av} are shown in figure \ref{fig:spactralAV}.
At every frequency channel separation, the SAV decreased as was integrated for longer time.
While the SAV was decreasing as a function of frequency channel separation at shorter integration than 16 s, it appeared a local bottom and a top for longer integration than 16 s.
They appeared at $< 200$ ch and $\sim 259$ ch, respectively.
Figure \ref{fig:SAV_fit} shows the SAV of a 64-min-integrated spectrum as an example.
The profile was composed of two power-law components.
For narrower channel separations than the bottom, the SAVs was dominated by a power-law profile with the index of $-2.003 \pm 0.001$.
The index above the bottom was $1.29 \pm 0.02$.
The bottom of the SAV appeared near the frequency span where the two power-law components were equivalent.

At longer time integration, the bottom shifted towards narrower spectral separation in longer integration time, while the top stayed at almost the same separation.
These behavior is summarized in table \ref{tab:spectralAV}.

\subsection{Time Stability} \label{subsec:timeVariability}
\begin{figure}[h!]
\begin{center}
\FigureFile(80mm,50mm){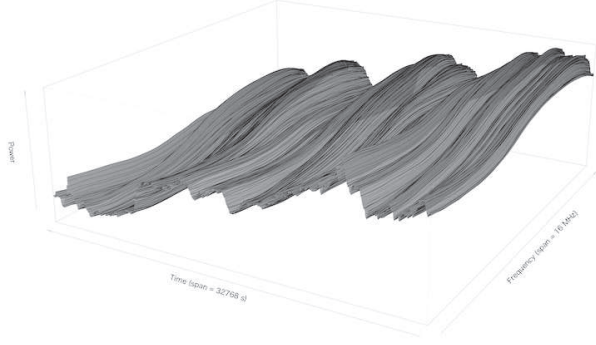}
\caption{Time variation of the bandpass-corrected spectrum, $\displaystyle \frac{\bar{H}_1(\nu, t)}{H_0(\nu)}$. Each spectrum was time-averaged for 30 s and spectral-smoothed with a 45-ch B-spline window.}
\label{fig:SpectralVariation}
\end{center}
\end{figure}

Figure \ref{fig:SpectralVariation} displays variability of $\displaystyle \frac{\bar{H}_1(\nu, t)}{\bar{H}_0(\nu)}$.
The smoothed spectrum shows time variation in terms of not only the power level but also its shape such as slope, curvature, and local bumps.
Variation of the bandpass was evaluated by the standard deviation of the smoothed spectrum, $\sigma_t(\nu)$, defined as $\displaystyle \sigma^2_t(\nu) = \left< \left(\frac{\bar{H}_1(\nu, t)}{\bar{H}_0(\nu)}\right)^2 \right> - \left< \frac{\bar{H}_1(\nu, t)}{\bar{H}_0(\nu)} \right>^2$.
Here, the expectation is taken by time average.

\begin{figure}[h!]
\begin{center}
\FigureFile(80mm,50mm){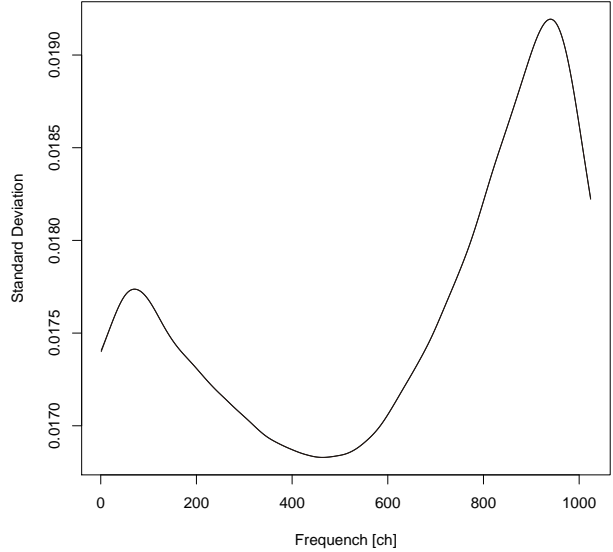}
\caption{Standard deviation (SD) of each spectral channel in the bandpass-corrected spectrum, $\displaystyle \sigma_t(\nu) = \sqrt{\left< \left(\frac{\bar{H}_1(\nu, t)}{H_0(\nu)}\right)^2 \right> - \left< \frac{\bar{H}_1(\nu, t)}{H_0(\nu)} \right>^2}$. The spectrum was time-averaged for 30 s and spectral-smoothed with a 45-ch B-spline window before calculation of the SD. Two peaks in the SD appeared at the frequency channels of 71 and 920.}
\label{fig:SpectralSD}
\end{center}
\end{figure}

Figure \ref{fig:SpectralSD} shows the standard deviation of the smoothed spectrum.
Two clear peaks appeared at 71 ch and 920 ch, near the shoulders of the bandpass, $H_0(\nu)$.
The standard deviation of $\sim 2$\% was significantly greater than the target accuracy of $\frac{\sigma}{T_{\mathrm{sys}}} = 5 \times 10^{-4}$.
Hence, it is necessary to calibrate the bandpass using the smoothed OFF-source spectrum within the timescale while $\displaystyle \frac{\bar{H}_1(\nu, t)}{\bar{H}_0(\nu)}$ is stable enough.

The timescale of stability was evaluated using the TAV defined in equation \ref{eqn:def_time_av}.
Since the top and bottom channels of the smoothed bandpass were also time variable, we instead used the twin peak channels in $\sigma_t(\nu)$, i.e.
we substituted $\Delta H_1(t) = H_1(71 {\rm ch}, t) - H_1(921 {\rm ch}, t)$.
The TAV is shown in figure \ref{fig:TimeAV}.
The first bottom of 30-sec-integrated spectrum appeared at 300 s.
The power-law index of the TAV was $\sim -2$ for $\tau < 60$ s and was $-1.7$ at $\tau = 80$, which we regulate as the spectral stability timescale in following processes.

\subsection{Scan Patterns and Performance}
Basing on above results we set the new scan pattern and compared with the conventional ON--OFF scans without spectral smoothing.
The conventional pattern (case 1) consists of pairs of 30-s ON and 30-s OFF scans.
The new scan pattern (case 2) was a set of 70-s ON and 10-s OFF with the SBC.
The pattern was designed as the cycle period to be shorter than the spectral stability timescale of 80 s and the optimal ON-to-OFF integration time ratio described in equation \ref{eqn:smoothed_sigma} with $N_{\mathrm{sw}} = 45$.

To reach the targeted noise level of $\sigma/T_{\mathrm{sys}} = 5 \times 10^{-4}$, the conventional pattern required 20 sets (1200 s) of the total telescope time.
Its resultant spectrum is shown in figure \ref{fig:resultSpectra} (A1).
The case-2 spectrum with the telescope time of 1200 s (15 sets of 80-s scans) is shown in figure \ref{fig:resultSpectra} (C1).
The SD was $(2.8^{+0.3}_{-0.2}) \times 10^{-4}$ (median, minimum, and maximum of 27 samples), $\times \frac{1}{1.74}$ that of the conventional scan pattern.
To achieve the targeted noise level, case 2 required 5 sets (400 s). Its resultant spectrum is shown in figure \ref{fig:resultSpectra} (B1).

Since the resultant spectrum appeared wiggled features other than random noise, we attempted baseline fitting and subtraction using the 3rd-order B-spline function with the node interval of 45 ch.
After the baseline subtraction, the SDs of cases 1, 2 (1200 s), and 2 (400 s) became $(4.7^{+0.3}_{-0.2}) \times 10^{-4}$, $(2.6^{+0.2}_{-0.2}) \times 10^{-4}$, and $(4.5^{+0.3}_{-0.3}) \times 10^{-4}$, respectively, and shown in figures \ref{fig:resultSpectra} (A2, C2, and B2).

The performance of SD as functions of total telescope time is summarized in table \ref{tab:rms_integ} and figure \ref{fig:residual_rms}.
At any total telescope time, the SBC exhibited better performance than conventional scans. While the SD of the conventional scans followed $\sigma \propto t^{-0.5}_{\mathrm{integ}}$ dependence, as was expected, the SBC resulted in a shallower slope with the power index of $-0.46$.

\subsection{Smoothing Window}
The optimal spectral smoothing window, $\Delta \nu$, was determined by the bottom in the SAV listed in table \ref{tab:spectralAV}.
For the timescale longer than 16 min, the bottom appeared 
at $\Delta \nu / \nu_{\mathrm{sp}} \sim 32 - 60$ ch.
We chose the window of 45 ch for time stability tests and spectral performance tests in previous subsections.
We also tested the spectral performance with various windows of 2, 3, 4, $\dots$, 191 ch for the scan pattern of case 2 with the total telescope time of 400 and 1200 s.

Figure \ref{fig:residual_SD} shows the results.
The median values of SDs in the resultant spectrum before baseline subtraction recorded a minimum at the 64-ch window for both telescope times of 400 s and 1200 s.
After baseline subtraction, the SD became significantly lower and the minimum appeared a flat bed for wider window than 45 ch.
In all cases the bottom was smaller than the targeted noise level.

\section{Discussion}

\subsection{Bandpass Flatness}
As is shown in figure 1, the SAV across bandwidth is composed of the thermal noise and fluctuation of bandpass characteristics.
The crossover point, where the thermal noise and the bandpass fluctuation is equivalent, appears at the bottom of the SAV.
Relevant smoothing window, $\Delta \nu$, should be set around the crossover point.
For longer integration the bottom of SAV shifts toward narrower channel separation.
The regression of the crossover point indicates that the channel separation of the bottom relates to the integration time as $\Delta \nu_{\mathrm{bottom}} \propto t^{-0.273\pm 0.005}_{\mathrm{integ}}$ for $t_{\mathrm{integ}} < 128$ min.

This behavior is explained as following consideration.
The thermal noise component is stochastic which does not depend on time and frequency.
Thus the SAV of thermal noise component $\propto \Delta \nu^{-2} t^{-1}_{\mathrm{integ}}$.
The bandpass fluctuation SAV $\propto \Delta \nu^{1.29 \pm 0.02}$.
This power-law index is slightly steeper than SAV $\propto \Delta \nu$ which is expected for a random-walk process along frequency.
Since the bandpass characteristics is frozen during short timescale, the SAV of bandpass fluctuation is independent of $t_{\mathrm{integ}}$.
Therefore, the crossover point appear under the condition of $\Delta \nu^{-2} t^{-1}_{\mathrm{integ}} = \alpha \Delta \nu^{1.29}$ where $\alpha$ is a proportional coefficient.
Thus, the SAV bottom will appear at $\Delta \nu_{\mathrm{bottom}} \propto t^{-\frac{1}{3.29}}_{\mathrm{integ}}$.
The behavior of the test observation, which indicates $\Delta \nu_{\mathrm{bottom}} \propto t^{-0.273\pm 0.005}_{\mathrm{integ}}$, is consistent with the theoretical consideration.

Although a different combination of a receiving system and a spectrometer other than the VERA system may have different properties of bandpass flatness, measuring SAVs at various integration time allows us to decompose it into thermal-noise and random-walk components.

\subsection{Expected Time Reduction}
As shown in subsection 3.4, the SBC (case 2) allows us to shorten the total telscope time to $\frac{1}{3}$ to achieve the targeted noise level, compared with the conventional method (case 1).
In other words, the SBC reduces the SD by a factor of $\sqrt{3}$ within the same total telescope time.
These results demonstrate that the telescope time efficiency can be tripled by the SBC method with the optimal scan pattern.

\begin{figure}[h!]
\begin{center}
\FigureFile(80mm,50mm){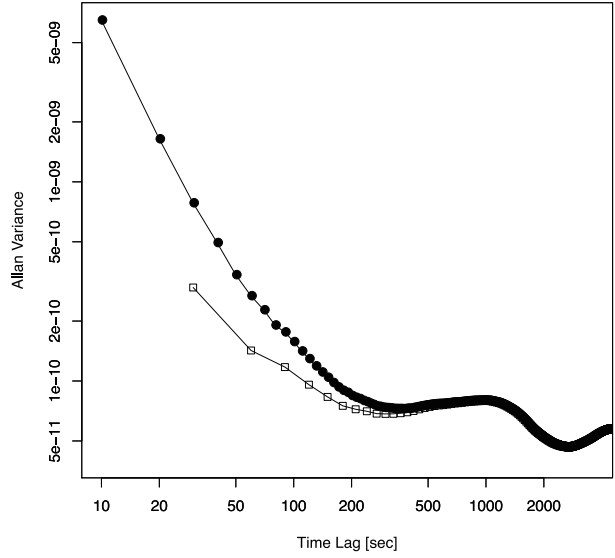}
\caption{Time-based Allan variance (TAV), $\sigma^2_{\mathrm{y}}(\tau)$, of spectral difference between 71 ch and 920 ch as a function of time lag as defined in equation \ref{eqn:def_time_av}. Sequences of filled circles and open squares represent 10-s and 30-s time-integrated spectra, respectively. The local TAV minima appeared at 370 s and 300 s for each. Significant departure from the power-law index of $-2$ arose at $\tau \ltsim 60$ s.}
\label{fig:TimeAV}
\end{center}
\end{figure}

\begin{figure}[h!]
\begin{center}
\FigureFile(160mm,100mm){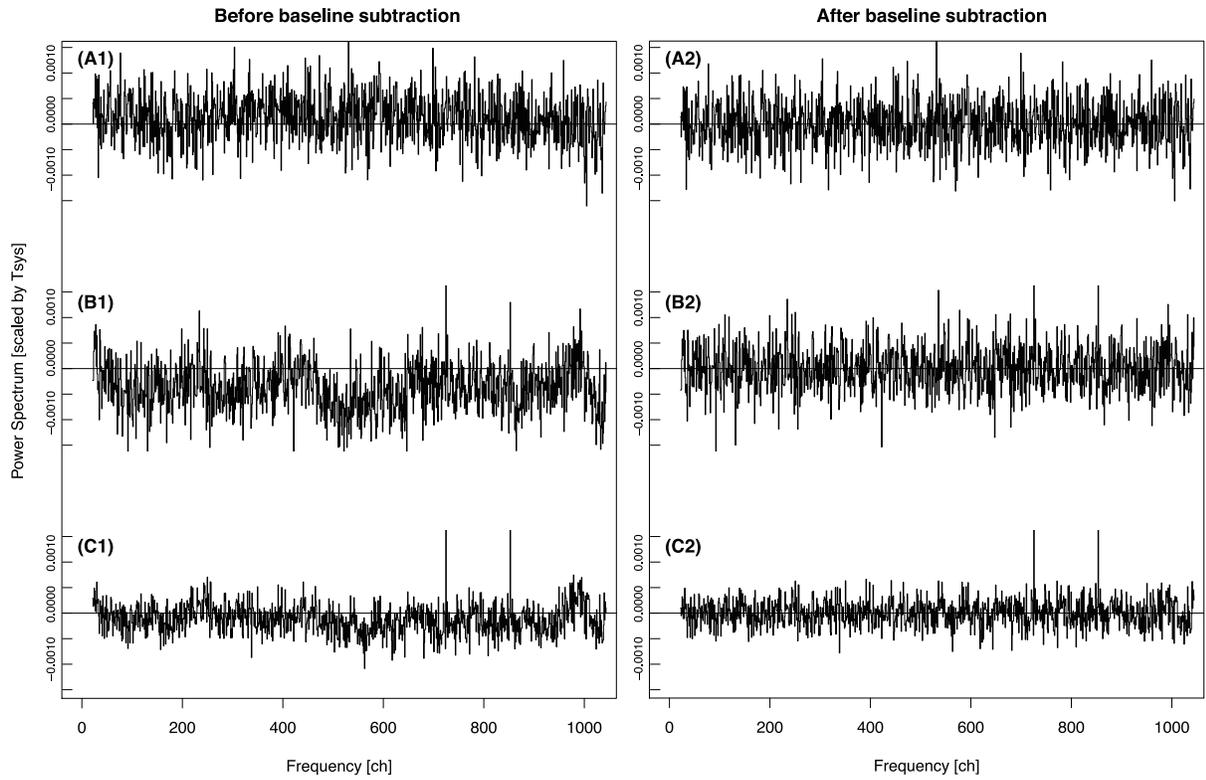}
\caption{Resultant spectra after bandpass correction. Left and right panels stand for before and after baseline subtraction performed after bandpass correction is applied. A1 and A2 : conventional ON--OFF scans (case 1) of the total telescope time of 1200 s. B1 and B2 : 70-sec On -- 10-s Off scans with spline smoothing (case 2) with the total telescope time of 400 s. C1 and C2 : case-2 scans with the telescope time of 1200 s.}
\label{fig:resultSpectra}
\end{center}
\end{figure}

\begin{figure}[h!]
\begin{center}
\FigureFile(80mm,50mm){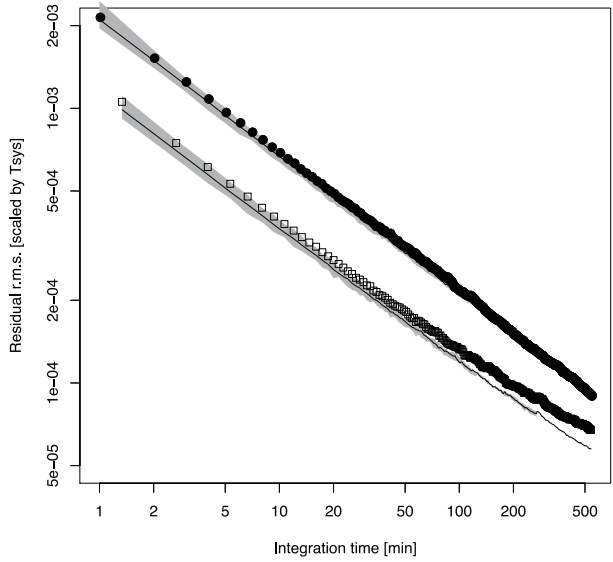}
\caption{SDs of the resultant spectra derived by ON--OFF scans as functions of total telescope time. Filled circles indicate results from sets of conventional 30-s ON + 30-s OFF scans without spectral smoothing. Opened squares represent those of 70-s ON + 10-s spline-smoothed OFF scans. The solid lines indicate median values of SDs obtained after baseline-subtraction was applied. The median was taken from a number of sets in 10-hour observations. The grey shade indicates maximum and minimum values of the sets. While the power-law index of the median values (solid lines) in conventional scans were $-0.50$ between 60 and 32400 s, that in proposed scan was $-0.46$ between 80 and 32400 s.}
\label{fig:residual_rms}
\end{center}
\end{figure}

\begin{figure}[h!]
\begin{center}
\FigureFile(80mm,50mm){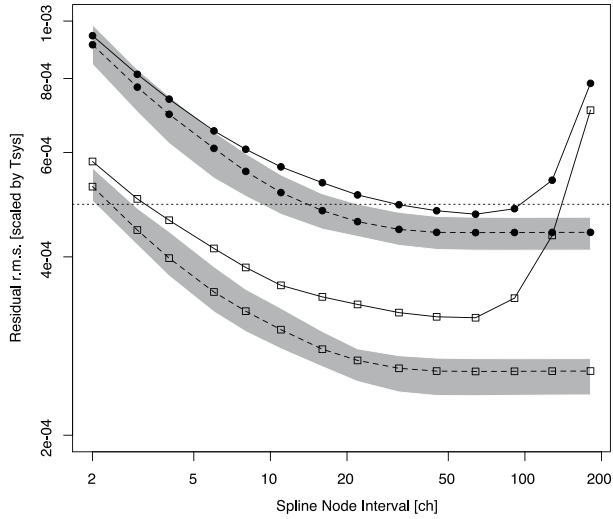}
\caption{Standard deviations (SDs) of the OFF-source-subtracted spectra. Filled circles and opened squares stand for scan patterns of 
case 2 with the telescope times of 400 s and 1200 s,
respectively. Solid and dashed lines indicate SDs before and after baseline correction with 45-ch spline smoothing is applied, respectively. The horizontal dotted line shows the SD level of case 1 (conventional ON--OFF scans without spectral smoothing). Median values are plotted with the symbols. The grey shades indicate the peak-to-peak range of the baseline-subtracted SDs.}
\label{fig:residual_SD}
\end{center}
\end{figure}

Since our analysis did not take scan gaps time into consideration, the efficiency can be different in realistic observations.
As shown in figure \ref{fig:resultSpectra}, systematic undulation remains in the case-2 spectrum before baseline subtraction.
The undulation can be ascribed to OFF-scan time interval of 70 s that may be longer than the stability timescale ({\bf see figure \ref{fig:TimeAV}}).
OFF-scan intervals should be shorter to reduce the undulation, however, this modification increases the time loss in scan gaps.
Optimal scan pattern should be tested in detail under realistic conditions of scan gaps and time stability. 

As shown in figure \ref{fig:residual_rms}, the slope of SD reduction by time integration for the SBC was shallower than that for conventional scans.
The shallower slope indicates that the SBC underperforms at a long integration time.
This underperformance can be caused by time instability of the bandpass.
The power index of the TAV of $-1.7$ at $\tau = 80$ s, with a small departure from $-2$, indicates that bandpass fluctuation remains with the level of $\sigma_{\mathrm{y}} \sim 10^{-5}$ ({\bf see figure \ref{fig:TimeAV}}).
This level of fluctuation is not tangible in the conventional scans even at the longest integration ($\frac{\sigma}{T_{\mathrm{sys}}} = 9.3 \times 10^{-5}$) but can affect in the SBC ($\frac{\sigma}{T_{\mathrm{sys}}} = 6.3 \times 10^{-5}$).
The bandpass stability can be the principal component to determine the sensitivity at a longer integration time then 32400 s.
To pursue the sensitivity at the longer integration, shorter scan pattern should be considered in spite of time loss in scan gaps.

\subsection{Advantages in a Dual Beam System}
Dual beam systems are equipped in some radio telescopes such as the GBT 100-m \citep{2002ursi.confE...3J}, the Nobeyama 45-m \citep{2010ASJM...V45a}, and the VERA \citep{2003ASPC..306..367K}, to double the time efficiency by interchanging ON- and OFF-source scans between two beams.
The duty cycle for ON and OFF scans in dual beam systems should be 1:1 (or $x=0.5$ in equation \ref{eqn:min_sigma}), if performance of two receivers is uniform, and different from the optimal duty cycle for the SBC.
In this case the variance, $\sigma^2$, will be reduced by a factor of $\displaystyle \frac{1}{2} \left(1+\frac{1}{N_{\mathrm{sw}}}\right)$ when we employ the SBC.
Taking $N_{\mathrm{sw}} = 45$ for instance, the variance is reduced by the factor of $0.51$ and the telescope time efficiency is almost doubled, 3.9 times better than conventional ON--OFF scans with a single beam system, or 1.3 times better than the SBC with a single beam system.

Consideration in this subsection implies that the SBC for single-beam with conventional ON--OFF scans (1:1 duty cycles) also doubles SNR.
Re-analysis with the SBC for previous observations can offer an opportunity to enhance SNR, if ON- and OFF-source spectra are separately recorded, and the efficiency depends on the spectral stability of the system.

\subsection{Advantages in OTF Scans}
ON-the-fly (OTF) mapping shares the OFF-source pointing among a continuous scan pass \citep{2007A&A...474..679M}.
\citet{2008PASJ...60..445S} argued that 
the noise level achieved in the unit observing time
of the OTF map is written as
\begin{eqnarray}
\Delta T^*_{\mathrm{A}}(0) &=&  \frac{T_{\mathrm{sys}}}{\sqrt{B}} \sqrt{\left( \frac{1}{t^{\mathrm{ON}}_{\mathrm{cell}}} + \frac{1}{t^{\mathrm{OFF}}_{\mathrm{cell}}} \right) } \nonumber \\
& \times & \sqrt{\left( t_{\mathrm{scan}}+t_{\mathrm{OH}} + \frac{t_{\mathrm{OFF}}}{N^{\mathrm{SEQ}}_{\mathrm{scan}}} \right) N_{\mathrm{row}} f_{\mathrm{cal}}}, \label{eqn:sawada2008} 
\end{eqnarray}
and derived the optimal OFF-source integration time to map a rectangular area with the row span of $l_1$ was given $\displaystyle t_{\mathrm{OFF}} = \sqrt{(t_{\mathrm{scan}} + t_{\mathrm{OH}}) \frac{\eta d t_{\mathrm{scan}}}{l_1}} \sqrt{N^{\mathrm{SEQ}}_{\mathrm{scan}}}$, where $t_{\mathrm{scan}} = l_1 / v_{\mathrm{scan}}$ is the ON-source scan time for a row by the scan speed of $v_{\mathrm{scan}}$, $t_{\mathrm{OH}}$ is the overhead time between row scans, $\eta$ is the efficiency determined by the gridding convolution function, $d$ is the grid spacing, and $N^{\mathrm{SEQ}}_{\mathrm{scan}}$ is the number of row scans between OFF-source integrations.
In the practical case of $\eta = 4.3$, $l_1 = 600^{\prime \prime}$, $\Delta l = 5^{\prime \prime}$, $d = 7^{\prime \prime}.5$, $t_{\mathrm{OH}} = 25$ s, and $N^{\mathrm{SEQ}}_{\mathrm{scan}} = 1$, according to \citet {2008PASJ...60..445S}, the ON-source scan time of $t_{\mathrm{scan}} =20$, $40$, and $60$ s yields $t^{\mathrm{optimal}}_{\mathrm{OFF}} = 7$, 12, and 17 s, respectively.

The optimization is modified when the SBC is applied.
Equation \ref{eqn:sawada2008} is modified as
\begin{eqnarray}
\Delta T^*_{\mathrm{A}}(0) &=&  \frac{T_{\mathrm{sys}}}{\sqrt{B}} \sqrt{\left( \frac{1}{t^{\mathrm{ON}}_{\mathrm{cell}}} + \frac{1}{N_{\mathrm{sw}} t^{\mathrm{OFF}}_{\mathrm{cell}}} \right) } \nonumber \\
& \times & \sqrt{\left( t_{\mathrm{scan}}+t_{\mathrm{OH}} + \frac{t_{\mathrm{OFF}}}{N^{\mathrm{SEQ}}_{\mathrm{scan}}} \right) N_{\mathrm{row}} f_{\mathrm{cal}}}, \label{eqn:OTF_BP} 
\end{eqnarray}
when the smoothing window is $N_{\mathrm{sw}}$ ch.
The optimal OFF-source integration will be $\displaystyle t_{\mathrm{OFF}} = \sqrt{(t_{\mathrm{scan}} + t_{\mathrm{OH}}) \frac{\eta d t_{\mathrm{scan}}}{N_{\mathrm{sw}} l_1}} \sqrt{N_{\mathrm{scan}}^{\mathrm{SEQ}}}$.
For comparison to \citet{2008PASJ...60..445S}, $t_{\mathrm{scan}} =20$, $40$, and $60$ s yields $t^{\mathrm{optimal}}_{\mathrm{OFF}} = 1$, 1.8, and 2.5 s when we take $N_{\mathrm{sw}} = 45$, and the noise level decreases by factors of 0.886, 0.874, and 0.856, respectively.
Thus, the total telescope time that achieves the same SD will be reduced by a factor of $\sim 1.3 - 1.4$.

The effect of the SBC is less than simple ON-OFF scans, nevertheless, the SBC offers somewhat better results.
Consideration in this subsection does not involve bandpass fluctuation.
More realistic optimization and noise level should be estimated basing on time-stability of receiving systems.

\subsection{Vulnerability against Spurious Signals}
SBC yields weakness against unwanted spurious emissions such as RFI, contamination of sampling clocks and reference signals in a receiving system, artificial pattern caused by numerical errors of spectral calculation, etc.
These spurious signals usually show spiky features in both ON- and OFF-scan spectra.
They are suppressed in the smoothed OFF-scan spectra while kept in the unsmoothed ON-scan spectra.
Thus they will appear in the final results, though they would be subtracted by the conventional ON - OFF scans if they were stable.  

In our test observations, spurious emissions appeared at 704 ch and 832 ch (see figure \ref{fig:resultSpectra}) that correspond to 11/16 and 13/16 of the bandwidth.
Their line width is narrower than the spectral resolution.
Digital noise such as harmonics of the sampling clock can cause such spurious emissions whose frequencies are simple fractions of the bandwidth.
These spurious frequencies should be masked in spectral data reduction.
Observers need to pay deeper attention to spurious when they apply SBC.

\section{Summary}
It is presented that the proposed SBC method offers significantly better performance than conventional ON--OFF scans, especially when a stable spectrometer such as digital system was equipped.
Our tests showed that the total telescope time was reduced to $\frac{1}{3}$ to attain the same SNR, or $\times 1.7$ better SNR was obtained in the same telescope time for a single-pointing ON-OFF scans.
The SBC method was also efficient for dual-beam systems, ON-the-fly mapping.

The optimal analysis scheme of the SBC is presented in this paper.
To apply this method, radio observatories should offer 

\begin{itemize}
\item Long-time integrated bandpass to obtain good-enough $H_0(\nu)$. This can be done in off seasons.

\item SAV to show flatness of $H_1(\nu, t)$ to allow observers to compute the optimal smoothing window.

\item TAV to estimate stability timescale of $H_1(\nu, t)$ to design the optimal scan pattern and total telescope time required.

\item Spectral outputs of ON- and OFF-source, separately. Automated ON-OFF subtraction makes it impossible to OFF-source spectral smoothing in post-observation analysis.

\end{itemize}

Finally, re-analysis of previous spectral data with the SBC can also offer an opportunity to enhance SNR, if ON- and OFF-source spectra are separately recorded.

\bigskip

The VERA Iriki 20-m telescope is a part of the VERA VLBI network which is operated by the National Astronomical Observatory of Japan and staffs and students of the Kagoshima University. 


\begin{table}
\caption{Bottom and top appeared in the spectral AV}
\label{tab:spectralAV}
\begin{center}
\begin{tabular}{rrrrr} \hline
Integ. & bottom & bottom SAV & top & top SAV \\
(min)  & (ch) & $\times 10^{-10}$ & (ch) &              $\times 10^{-10}$\\ \hline
  1 & 128 & $5.3$ & 255 & $7.2$ \\
  2 & 103 & $4.0$ & 257 & $7.0$ \\
  4 &  86 & $3.1$ & 259 & $6.8$ \\
  8 &  71 & $2.4$ & 259 & $6.7$ \\
 16 &  60 & $1.9$ & 259 & $6.7$ \\
 32 &  49 & $1.4$ & 259 & $6.5$ \\
 64 &  42 & $1.1$ & 259 & $6.5$ \\
128 &  39 & $0.9$ & 259 & $6.4$ \\
256 &  34 & $0.7$ & 259 & $6.3$ \\
512 &  32 & $0.6$ & 259 & $6.3$ \\ \hline
\end{tabular}
\end{center}
\end{table}

\begin{table}
\caption{Spectral performance as functions of total telescope time}
\label{tab:rms_integ}
\begin{center}
\begin{tabular}{rllll} \hline
$t_{\mathrm{tot}}$ & \multicolumn{4}{c}{$\sigma / T_{\mathrm{sys}}$ ($\times 10^{-4}$)} \\ \cline{2-5}
 (min)       &  case 1a  & case 1b & case 2a & case 2b  \\ \hline
  4 & $10.8^{+3.7}_{-0.8}$    & $10.8^{+0.6}_{-0.7}$       & $6.1^{+0.5}_{-0.4}$    & $5.7^{+0.5}_{-0.4}$ \\
  8 & $~7.7^{+2.2}_{-0.5}$    & $~7.5^{+0.4}_{-0.4}$       & $4.3^{+0.4}_{-0.3}$    & $4.1^{+0.3}_{-0.2}$ \\
 16 & $~5.4^{+1.1}_{-0.4}$    & $~5.3^{+0.3}_{-0.3}$       & $3.1^{+0.3}_{-0.2}$    & $2.9^{+0.2}_{-0.1}$ \\
 32 & $~3.88^{+0.28}_{-0.25}$ & $~3.75^{+0.16}_{-0.20}$    & $2.25^{+0.11}_{-0.08}$ & $2.07^{+0.11}_{-0.10}$ \\
 64 & $~2.76^{+0.03}_{-0.14}$ & $~2.66^{+0.04}_{-0.10}$    & $1.63^{+0.05}_{-0.11}$ & $1.49^{+0.06}_{-0.05}$ \\
128 & $~1.93^{+0.05}_{-0.03}$ & $~1.88^{+0.006}_{-0.016}$  & $1.19^{+0.05}_{-0.04}$ & $1.07^{+0.02}_{-0.01}$ \\
256 & $~1.33^{+0.01}_{-0.01}$ & $~1.29^{+0.002}_{-0.003}$  & $0.89^{+0.01}_{-0.01}$ & $0.78^{+0.02}_{-0.02}$ \\
512 & $~0.93$                 & $~0.91$                    & $0.70$                 & $0.59$ \\ \hline
\multicolumn{5}{@{}l@{}}{\hbox to 0pt{\parbox{85mm}{\footnotesize Cases with 'a' and 'b' stand for before and after baseline subtraction, respectively.
\par\noindent
}\hss}}
\end{tabular}
\end{center}
\end{table}
%

%

\end{document}